\documentclass[12pt]{article}

\begin{document}

\title{ Exact solution to Fick-Jacobs equation}

\author{ Juan M. Romero\thanks{jromero@correo.cua.uam.mx},   O. Gonz\'alez-Gaxiola\thanks{ogonzalez@correo.cua.uam.mx} ,
 and G. Chac\'on-Acosta\thanks{gchacon@correo.cua.uam.mx}\\
 [0.5cm]
\it Departamento de Matem\'aticas Aplicadas y Sistemas,\\
\it Universidad Aut\'onoma Metropolitana-Cuajimalpa\\
\it M\'exico, D.F  01120, M\'exico\\[0.3cm]} %\\

\date{}

\pagestyle{plain}

\maketitle

\begin{abstract}
 A relation between   Fick-Jacobs  and   Schr\"{o}dinger equation is shown.
When  the  diffusion coefficient is constant, exact solutions for Fick-Jacobs equation are obtained. 
Using a change of variable the general case is studied.
\end{abstract}
.

\section{Introduction}
\label{s:Intro}

\noindent Recently,  methods of quantum physics haven been used to solve problems in other
disciplines. For instance, Black-Scholes can be mapped to  Schr\"{o}dinger equation  \cite{baaquie:gnus,jana:gnus,me:gnus}. It is worth to mention that Black-Scholes
equation  plays an important role in finance \cite{baaquie:gnus,jana:gnus,me:gnus}. Moreover,  
it is well known that the Fick  equation represents the  simplest model of diffussion. The Fick equation  
can be mapped to  Schr\"{o}dinger equation too.  When the diffusion is in a  channel which has shape of  surface of revolution with cross section of area $A(x),$  the Fick equation is changed to  Fick-Jacobs  equation \cite{jacobs:gnus}
\begin{eqnarray}
\frac{ \partial C(x,t)}{\partial t}&=&\frac{\partial }{\partial x}
\left[D(x)A(x) \frac{\partial }{\partial x} \left(
\frac{C(x,t)}{A(x)} \right)\right],\label{eq:fick0} 
\end{eqnarray}
where $C(x,t)$ is the concentration of particles and $D(x)$ is the diffusion coefficient.\\

\noindent The Fick-Jacobs equation is important to study diffusion in biological channels or zeolites 
and is studied with perturbative and numerical methods \cite{reguera:gnus,reguera1:gnus,reguera2:gnus,dagdug:gnus,dagdug1:gnus,zwanzig:gnus,zwanzig1:gnus}.  
In this paper we find a relation between Fick-Jacobs  and   Schr\"{o}dinger equation. Also, 
using quantum mechanics methods, we  obtain exact solutions for Fick-Jacobs equation, in particular 
when  the  diffusion coefficient is constant and  the shape of channel is  conical, throat-like, 
sinusoidal or Gaussian. To study the Fick-Jacobs  equation with constant diffusion coefficient, we use the operator 
\begin{eqnarray} 
 \hat P_{(f)}= -i \frac{\partial }{\partial x } + i\frac{\partial f( x) }{\partial x }, \qquad \quad x\in {\bf R} 
\label{eq:momentum0}
\end{eqnarray}
where $f$ is smooth function. The operator (\ref{eq:momentum0}) was first studied by P. A. M. Dirac \cite{Dirac:gnus}. The general 
case is studied with a change of variable.\\

\noindent This paper is organized as follows: In section 2 a review of quantum mechanics with a generalized operator (\ref{eq:momentum0}) is presented. In section 3 is 
shown a relation between  Fick-Jacobs and Schr\"{o}dinger equation. In section 4 some properties of solutions are   considered. In section 5 the case with constant diffusion coefficient is studied for different examples. 
In section 6 the general case is analyzed. Finally in section 7 a summary is given.

\section{ Generalized momentum operator}

\noindent The  Hermitian operator
\begin{eqnarray}
\hat P=-i \frac{\partial }{\partial x }
\end{eqnarray}
is the usual  momentum operator in quantum
mechanics, with $\hbar=1$. However, in early quantum mechanics, P. A. M.
Dirac studied the no-Hermitean operator \cite{Dirac:gnus}

\begin{eqnarray}
 \hat P_{(f)} = e^{f(x)} \hat P e^{-f(x)}= \hat P + i\frac{\partial f( x) }{\partial x }.  
 \label{eq:momentum}
\end{eqnarray}
This operator is very interesting, for example using it  we can build  a supersymmetric
quantum mechanics. In fact, lets us propose the Hermitian Hamiltonians:
\begin{eqnarray}
& & \hat H_{1}=\alpha^{2} \hat P_{(f)}^{\dagger} \hat P_{(f)} 
= \alpha^{2}\left( \hat P^{\;2}+\frac{d^{2}f}{dx^{2}} +\left( \frac{df}{dx} \right)^{2} \right),\label{eq:h1} \\
& & \hat H_{2}= \alpha^{2} \hat P_{(f)} \hat
P_{(f)}^{\dagger}=\alpha^{2}\left(\hat P^{\;2}
-\frac{d^{2}f}{dx^{2}}+ \left( \frac{df}{dx} \right)^{2}  \right),
\label{eq:h2}
\end{eqnarray}
here $ \alpha$ is constant. Now,  if $W$ is a smooth function, we can propose 
\begin{eqnarray}
f(x)=\int_{0}^{x}W(u)du, \qquad x\geq 0,
\end{eqnarray}
then
\begin{eqnarray}
\hat H_{1}&=&\alpha^{2}\left( \hat P^{\;2}+\frac{dW}{dx} +W^{2} \right), \\
\hat H_{2}&=& \alpha^{2}\left(\hat P^{\;2}-\frac{dW}{dx}+ W^{2}
\right).
\end{eqnarray}
These  Hamiltonians can be used to form the matrix
\begin{eqnarray}
\hat {\mathcal H}&=&\left(
\begin{array}{rr}
 \hat H_{1} & 0\\
 0&  \hat H_{2}
\end{array}
\right). 
\end{eqnarray}
According to the supersymmetric quantum mechanics \cite{cooper:gnus}, $\hat {\mathcal H}$  represents a superhamiltonian. \\

\noindent Furthermore, with the  operator (\ref{eq:momentum}) we can  build
the Hamiltonians:
\begin{eqnarray}
& & \hat H_{3}=\beta^{2}  \hat P_{(f)}^{\dagger} \hat P_{(f)}^{\dagger}= \beta^{2}\left( \hat P^{2} -2i\frac{df}{dx}\hat P
-\frac{d^{2}f}{dx^{2}} -\left(\frac{df}{dx}  \right)^{2}   \right),\label{eq:3}\\
& & \hat H_{4} =\beta^{2}  \hat P_{(f)} \hat P_{(f)}=\beta^{2}\left(
\hat P^{2} +2i \frac{df}{dx} \hat P
+\frac{d^{2}f}{dx^{2}}-\left(\frac{df}{dx}\right)^{2}
\right)\label{eq:4}, 
\end{eqnarray}
where $\beta$ is constant.  These Hamiltonians  are no-Hermitian, however with them  we can construct Hamiltonians which arise  
in the so-called quantum finance \cite{me:gnus}.  We will use  these Hamiltonians  to study 
 Fick-Jacobs equation.\\

\noindent Now, using the Hamiltonians $\hat H_{3}$ and $\hat H_{4},$ we can set the  wave equations
\begin{eqnarray}
i\frac{\partial \psi_{3}(x,t)}{\partial t}&=&\hat H_{3}\psi_{3}(x,t),  \label{eq:ham1}\\
i\frac{\partial \psi_{4}(x,t)}{\partial t}&=&\hat
H_{4}\psi_{4}(x,t).   \label{eq:ham2}
\end{eqnarray}
Both of these equtions are  equivalent to non-relativistic free particle. In fact, first note that
\begin{eqnarray}
\hat P_{(f)}^{\dagger} \hat P_{(f)}^{\dagger}&=& e^{-f(x)}\hat P e^{f(x)} e^{-f(x)}\hat Pe^{f(x)} = e^{-f(x)}\hat P^{2}e^{f(x)},\nonumber\\
\hat P_{(f)} \hat P_{(f)}&=& e^{f(x)}\hat Pe^{-f(x)} e^{f(x)}\hat
Pe^{-f(x)} = e^{f(x)}\hat P^{2}e^{-f(x)},
\end{eqnarray}
namely 
\begin{eqnarray}
i\frac{\partial \psi_{3}(x,t)}{\partial t}&=&\hat H_{3}\psi_{3}(x,t) = \beta^{2} e^{-f(x)}\hat P^{2}e^{f(x)}\psi_{3}(x,t), \nonumber\\
i\frac{\partial \psi_{4}(x,t)}{\partial t}&=&\hat H_{4}\psi_{4}(x,t) = \beta^{2} e^{f(x)}\hat P^{2}e^{-f(x)}\psi_{4}(x,t), \nonumber
\end{eqnarray}
thus
\begin{eqnarray}
i\frac{\partial \left( e^{f(x)} \psi_{3}(x,t)\right) }{\partial t}&=& \beta^{2} \hat P^{2} \left(e^{f(x)} \psi_{3}(x,t)\right) , \nonumber \\
i\frac{\partial  \left(e^{-f(x)} \psi_{4}(x,t)\right)}{\partial t}&=& \beta^{2} \hat P^{2} \left(e^{-f(x)} \psi_{4}(x,t)\right). \nonumber
\end{eqnarray}
These  equations represent the free particle wave equation and their  solutions are:
\begin{eqnarray}
 e^{f(x)} \psi_{3}(x,t)= e^{-i\beta^{2} \hat P^{2}t} \left(e^{f(x)} \psi_{03}(x)\right) , \nonumber\\
  e^{-f(x)} \psi_{4}(x,t)= e^{-i\beta^{2} \hat P^{2}t} \left(e^{-f(x)} \psi_{04}(x)\right),
\end{eqnarray}
where $\psi_{3}(x,0)= \psi_{03}(x)$ and  $\psi_{4}(x,0)= \psi_{04}(x)$
are the initial conditions. Therefore, 
\begin{eqnarray}
 \psi_{3}(x,t)= \left( e^{-f(x)} e^{-i\beta^{2} \hat P^{2}t} e^{f(x)}\right) \psi_{03}(x) , \nonumber\\
\psi_{4}(x,t)= \left( e^{f(x)} e^{-i\beta^{2} \hat P^{2}t} e^{-f(x)} \right) \psi_{04}(x)
\end{eqnarray}
are the general solution of equations  (\ref{eq:ham1}) and (\ref{eq:ham2}).\\

\noindent Also, if $V$ is a potential, we can prove that the wave equations 
\begin{eqnarray}
i\frac{\partial \psi_{3}(x,t)}{\partial t}&=&\left( \beta^{2}\hat P_{(f)}^{\dagger} \hat P_{(f)}^{\dagger}+V(x)  \right)\psi_{3}(x,t) \label{eq:ham3}\\
i\frac{\partial \psi_{4}(x,t)}{\partial t}&=&
\left(\beta^{2} \hat P_{(f)} \hat P_{(f)}+V(x)\right) \psi_{4}(x,t), \label{eq:ham4}
\end{eqnarray}
are equivalent to 
\begin{eqnarray}
i\frac{\partial \left( e^{f(x)}\psi_{3}(x,t)\right)}{\partial t}&=&\hat H \left( e^{f(x)}\psi_{3}(x,t)\right),
\label{eq:ham5} \\
i\frac{\partial \left( e^{-f(x)}\psi_{4}(x,t)\right)}{\partial t}&=&\hat H \left( e^{-f(x)}\psi_{4}(x,t)\right),
\label{eq:ham6}
\end{eqnarray}
here the Hamiltonian is 
\begin{eqnarray}
\hat H= \beta^{2} \hat P^{2}+V(x).
\end{eqnarray}
The general solutions of the equations (\ref{eq:ham5}) and
(\ref{eq:ham6}) are:
\begin{eqnarray}
e^{f(x)}\psi_{3}(x,t)&=&e^{-i\hat Ht} \left( e^{f(x)}\psi_{03}(x)\right),\\
 e^{-f(x)}\psi_{4}(x,t)&=&e^{-i\hat Ht} \left( e^{-f(x)}\psi_{04}(x)\right),
\end{eqnarray}
therefore, the general solutions of the equations (\ref{eq:ham3})
and (\ref{eq:ham4}) are given by
\begin{eqnarray}
\psi_{3}(x,t)&=&\left( e^{-f(x)} e^{-i\hat Ht}  e^{f(x)} \right)\psi_{03}(x), \\
\psi_{4}(x,t)&=& \left( e^{f(x)}e^{-i\hat Ht}  e^{-f(x)} \right) \psi_{04}(x),
\end{eqnarray}
where the initial conditions are $\psi_{3}(x,0)= \psi_{03}(x)$ and  $\psi_{4}(x,0)= \psi_{04}(x).$\\

\noindent In the next section we will use these results to show a equivalence  between  
Fick-Jacobs  and  Schr\"{o}dinger equation.

\section{ Fick-Jacobs equation as  Schr\"{o}dinger equation   }

\noindent The Fick-Jacobs equation (\ref{eq:fick0}) can be written as
\begin{eqnarray}
\frac{ \partial C(x,t)}{\partial t}&=&D(x) \frac{\partial^{2} C(x,t) }{\partial x^{2}} +D(x)\frac{ \partial }{\partial x}\left[
\ln\left( \frac{D(x)}{A(x)}\right)\right]\frac{\partial C(x,t) }{\partial x}\nonumber\\
& & -\frac{ \partial }{\partial x}  \left( D(x) \frac{ \partial \ln A(x)}{\partial x} \right)C(x,t).\label{eq:fick}
\end{eqnarray}
Now, with the  change of variable
\begin{eqnarray}
y=\int_{x_0}^{x} \frac{dz}{\sqrt{D(z)}}, \qquad x_0={\rm constant},
\label{eq:fick2}
\end{eqnarray}
we have
\begin{eqnarray}
& &\frac{\partial }{\partial x}= \frac{1}{\sqrt{D(x)}} \frac{\partial
}{\partial y}, \quad D(x)\frac{\partial^{2} }{\partial x^{2}}=
\frac{\partial^{2} }{\partial y^{2}}-\frac{\partial
\left(\sqrt{D(x)}\right)}{\partial x} \frac{\partial }{\partial y}.\label{eq:change1}
\end{eqnarray}
Using (\ref{eq:change1})  in  (\ref{eq:fick}),  we get
\begin{eqnarray}
\frac{ \partial C(y,t)}{\partial t}&=&\frac{\partial^{2}C(y,t) }{\partial y^{2}} +
\sqrt{D(x)}\frac{ \partial }{\partial x}  \left(\ln\left[ \frac{\sqrt{D(x)}}{A(x)}\right]\right)
\frac{\partial C(y,t)}{\partial y} \nonumber\\
& &-  \frac{ \partial }{\partial x}  \left( D(x)\frac{\partial \ln A(x)}{\partial x}\right) C(y,t).
\label{eq:change}
\end{eqnarray}
Whether   the change of variable (\ref{eq:fick2}) is invertible,  this equation can be written as a function of 
variable $y$.  Also, if  (\ref{eq:fick2}) is invertible,  we can define the  functions:
\begin{eqnarray}
\frac{\partial}{\partial y} f(y) &=& -\frac{1}{2}
\sqrt{D(x)}\frac{ \partial }{\partial x}  \left[\ln\left( \frac{\sqrt{D(x)}}{A(x)}\right)\right] ,\nonumber\\
V(y)&=& \left(\frac{\partial f(y) }{\partial y}\right)^{2}-
\frac{\partial^{2} f(y) }{\partial y^{2}} +\frac{ \partial
}{\partial x}  \left(  D(x)\frac{\partial \ln A(x)}{\partial
x}\right).
\end{eqnarray}
Then (\ref{eq:change}) can be written as
\begin{eqnarray}
-\frac{ \partial C(y,t)}{\partial t}&=&\hat H_{f}  C(y,t), \label{eq:fick-1}
\end{eqnarray}
here 
\begin{eqnarray}
\hat H_{f}= \hat P^{2}+ 2i\frac{\partial f(y)}{\partial y} \hat P
+\frac{\partial^{2} f(y) }{\partial y^{2}} - \left(\frac{\partial
f(y) }{\partial y}\right)^{2}+ V(y),\quad \hat P=-i\frac{\partial
}{\partial y}.\nonumber
\end{eqnarray}
Moreover, using the results of Section 2, we obtain
\begin{eqnarray}
\hat H_{f}=e^{f(y)}\hat H e^{-f(y)}, \qquad \hat H=   \hat P^{2}+V(y),
\end{eqnarray}
then 
\begin{eqnarray}
-\frac{ \partial C(y,t)}{\partial t}&=& e^{f(y)}\hat H e^{-f(y)}  C(y,t),\label{eq:fick-2}
\end{eqnarray}
namely
\begin{eqnarray}
-\frac{ \partial\left( e^{-f(y)} C(y,t) \right)}{\partial t}&=& \hat H\left( e^{-f(y)}  C(y,t)\right)
\end{eqnarray}
and the  solution of this equation is given by 
\begin{eqnarray}
 e^{-f(y)} C(y,t)=e^{-\hat Ht}\left( e^{-f(y)}  C_{0}(y)\right).
\end{eqnarray}

Thus, when the change of variables (\ref{eq:fick2}) is
invertible, the general solution of  Fick-Jacobs equation is
\begin{eqnarray}
C(y,t)=\left( e^{f(y)}e^{-\hat Ht} e^{-f(y)} \right) C_{0}(y),
\label{eq:general}
\end{eqnarray}
with the initial condition $C(y,0)= C_{0}(y)$.\\

\noindent There are several examples where
the change of variable (\ref{eq:fick2}) is invertible, for instance $D={\rm constant}$, $D(x)=x^{n},
n=1,2,\cdots ,$ or $D(x)=e^{\alpha x}$, $\alpha\in \bf {R}$.\\

\section{Some properties of solutions}

\noindent Now  we will study some solutions properties of Fick-Jacobs equation.\\

\noindent First, lets us  take 
\begin{eqnarray}
C_{0}(y)= e^{f(y)} ,
\end{eqnarray}
then the solution (\ref{eq:general}) gives 
\begin{eqnarray}
C(y,t)=e^{f(y)}.
\end{eqnarray}
Namely, in this case the system does not evolve. It is a interesting result, because in this case there is not diffusion.\\

\noindent Now, whether  we want  to find the evolution of any condition initial, we can use quantum mechanics  
results. For example, when   the Hamiltonian is  
\begin{eqnarray}
\hat{H} = \hat{P}^2,
\end{eqnarray}
and   initial condition is 
\begin{eqnarray}
C_{0}(y) =  e^{f(y)} \frac{1}{(2 \pi
\sigma^2)^{1/2}}\,e^{-\frac{(y-a_0)^2}{4\sigma^2}},\qquad  a_{0}, \sigma={\rm constant}, 
\end{eqnarray}
using quantum mechanics results, we get 
\begin{eqnarray}
C(y,t) =  e^{f(y)}\left(\frac{\sigma^2}{2\pi}\right)^{1/4}\frac{1}{\left(\sigma^2+t\right)^{1/2}}\,e^{-\frac{(y-a_0)^2}{4\left(\sigma^2+t\right)}}.
\end{eqnarray}

\noindent In general, if  $\hat H\psi_n(y)=E_{n} \psi_n(y)$  and the concentration initial is 
\begin{eqnarray}
C_0(y)=  e^{f(y)}\phi_0(y).
\end{eqnarray}
with 
\begin{eqnarray}
\phi_0(y)=\sum_{n\geq0} a_n \psi_n(y),
\end{eqnarray}
where  $a_n$ are  Fourier coefficients,
then
\begin{eqnarray}
C(y,t)&=& \sum_{n\geq0} a_n e^{f(y)} e^{-E_n t}\psi_n(x).\nonumber 
\end{eqnarray}

\noindent Thus, as the Fick equation, the solution of  Fick-Jacobs equation can be seen as quantum states.  

\section{Fick-Jacobs equation with constant diffusion coefficient}

\noindent In this section we study the particular case $D=D_{0}=$ constant.\\

\noindent For this case the general solution  is given by
\begin{eqnarray}
C(x,t)=\sqrt{A(x)} e^{-\hat H t}
\left(\frac{C_{0}(x)}{\sqrt{A(x)}}\right), \label{eq:soln-gral}
\end{eqnarray}
with 
\begin{eqnarray}
& &  \hat H= D_{0} \left(\hat
P^{2}+V(x)\right), \qquad \hat P=-i \frac{\partial }{\partial x}, \\
& &V(x)=\frac{1}{2} \frac{ 1}{A(x)}\frac{\partial^{2} A(x)}{\partial x^{2}}
-\frac{1}{4}  \left( \frac{\partial \ln A(x)}{\partial x}\right)^{2}.
\label{eq:potencial0}
\end{eqnarray}
Now  lets us consider  the following 
geometries:

\subsection{Canonical channel}

When  the channel is a cone, the cross-sectional area is $A(x)=\pi
(1+\lambda x)^{2}$, where $\lambda$ is the slope of the generatriz of the cone.  For this case,   the
potential  (\ref{eq:potencial0}) is
\begin{eqnarray}
V(x)=0.
\end{eqnarray}
Then, using the results of Section 3, we find
\begin{eqnarray}
C(x,t)=(1+\lambda x) e^{-D_{0}\hat P^{2}}\left(\frac{C_{0}(x)}{1+\lambda x}\right),
\end{eqnarray}
here the initial condition $C(x,0)=C_{0}(x)$ is satisfied. \\

\noindent In particular case 
\begin{eqnarray}
C_0(x) = \frac{\left(1+\lambda x\right)}{(2 \pi
\sigma^2)^{1/2}}\,e^{-\frac{(x-a_0)^2}{4\sigma^2}},
\end{eqnarray}
we get
\begin{eqnarray}
C(x,t) = \left(1 + \lambda x\right)
\left(\frac{\sigma^2}{2\pi}\right)^{1/4}\frac{1}{\left(\sigma^2+tD_0\right)^{1/2}}\,e^{-\frac{(x-a_0)^2}{4\left(\sigma^2+tD_0\right)}}.
\end{eqnarray}

\noindent The conical tube was  solved in
\cite{dagdug:gnus} with a particular initial 
condition, here we solve the equation in general.\\

\subsection{Throat-like channel}

If the channel is  a
throat-like, the cross-sectional area can be taken as an exponential
function
\begin{eqnarray}
A(x)=e^{\alpha x+\beta},
\end{eqnarray}
where $\alpha$ and $\beta$ are  constants. In this case the
potential (\ref{eq:potencial0}) is
\begin{eqnarray}
V(x)=\frac{\alpha^{2}}{4}.
\end{eqnarray}
Therefore, the general solution is
\begin{eqnarray}\label{eq:sol-cte}
C(x,t)=e^{-\frac{D_{0}\alpha^{2} t}{4}} e^{\frac{\alpha x}{2}}
e^{-D_{0}\hat P^{2} t} e^{-\frac{\alpha x}{2}} C_{0}(x).
\end{eqnarray}
Then, when  the initial condition is a Gaussian
\begin{eqnarray}
C_0(x) = \frac{e^{\frac{\alpha}{2} x}}{(2 \pi
\sigma^2)^{1/2}}\,e^{-\frac{(x-a_0)^2}{4\sigma^2}},
\end{eqnarray}
  the concentration for any time is
\begin{eqnarray}
& &C(x,t) = e^{-\frac{D_0 \alpha^2t}{4}}\,e^{\frac{\alpha x}{2}}\,
\left(\frac{\sigma^2}{2\pi}\right)^{1/4}\frac{1}{\left(\sigma^2+tD_0\right)^{1/2}}\,e^{-\frac{(x-a_0)^2}{4\left(\sigma^2+tD_0\right)}}.
\end{eqnarray}

\subsection{Sinusoidal channel}

If the channel is sinusoidal,   the cross section is
\begin{eqnarray}
A(x)=B\left(\sin\gamma x\right)^2, \qquad B, \gamma={\rm constants}.
\end{eqnarray}
 In this case the potential  (\ref{eq:potencial0}) is 
\begin{eqnarray}
V(x)=-\gamma^2.
\end{eqnarray}
Then the solution is
\begin{eqnarray}
C(x,t)=e^{D_{0}\gamma^{2} t} \sin(\gamma x) e^{-D_{0}\hat P^{2} t}
\frac{ C_{0}(x)}{ \sin(\gamma x)}.
\end{eqnarray}
When condition  initial a Gaussian
\begin{eqnarray}
C_0(x) = \frac{\sin(\gamma x)}{(2 \pi
\sigma^2)^{1/2}}\,e^{-\frac{(x-a_0)^2}{4\sigma^2}},
\end{eqnarray}
the concentration for any time  is
\begin{eqnarray}
& & C(x,t) = e^{D_0 \gamma^2t}\sin(\gamma x)
\left(\frac{\sigma^2}{2\pi}\right)^{1/4}\frac{1}{\left(\sigma^2+tD_0\right)^{1/2}}\,e^{-\frac{(x-a_0)^2}{4\left(\sigma^2+tD_0\right)}}.
\end{eqnarray}

\subsection{Gaussian channel}

Whether the area of the cross section of the channel is a Gaussian, we have
\begin{eqnarray}
A(x)=e^{a x^{2}+bx+c},
\end{eqnarray}
here $a,b$ and $c$ are constants. In this case the potential defined in
(\ref{eq:potencial0})  is
\begin{eqnarray}
V(\zeta)=a^{2}\zeta^{2}+a,\quad {\rm with }\quad\zeta=x+\frac{b}{a},
\end{eqnarray}
and  the general solution is
\begin{eqnarray}
C(\zeta,t)=e^{-D_{0} at} e^{\frac{a\zeta ^{2}}{2}}
e^{-\hat h t} e^{-\frac{a\zeta
^{2}}{2}} C_{0}(\zeta),
\end{eqnarray}
where 
\begin{eqnarray}
\hat h=D_{0}\left(\hat P^{2}+a^{2}\zeta
^{2}\right),\label{eq:Ham-osc}
\end{eqnarray}
When we define   $m=1/(2D_{0})$ and $\omega^{2}=4D_{0}^{2}a^{2}$, the operator  (\ref{eq:Ham-osc})
is the Hamiltonian  oscillator harmonic.  Then, if  $\psi_{n}(\zeta)$ is eigenfunction of $\hat h,$ we have   
\begin{eqnarray}
 \hat h \psi_{n}(\zeta)= \psi_{n}(\zeta)E_{n},   \quad E_{n} = 2D_0a\left(n+\frac{1}{2}\right), \quad n=0,1,2,\cdots .\nonumber
\end{eqnarray}
Therefore, if  the initial condition is
\begin{eqnarray}
C_0(\zeta) =
e^{\frac{1}{2}a\zeta^2}\left(\frac{a}{\pi}\right)^{1/4}\psi_n(\zeta),
\end{eqnarray}
the concentration for any time is 
\begin{eqnarray}
C(\zeta,t) = 
e^{-D_0at}\,e^{\frac{a\zeta^2}{2}}\,e^{-2D_0a\left(n+\frac{1}{2}\right)t}\psi_n(\zeta).
\end{eqnarray}

\section{Other  cases}

\noindent In some  cases the  diffusion coefficient is given by \cite{reguera:gnus,zwanzig:gnus}
\begin{eqnarray}\label{eq:diffc}
D(x) =
\frac{D_0}{\left(1+\left[\frac{d}{dx}\left({\sqrt{\frac{A(x)}{\pi}}}\right)\right]^2
\right)^{1/2}}.
\end{eqnarray}
When the  change of variable (\ref{eq:fick2}) is invertible, we can study these cases. 
For instance, for  the conical channel, which has  cross section
$A(x)=\pi (1+\lambda x)^2$, the change of variable in this case is
$$y = x \left(\frac{\sqrt{1+\lambda^2}}{D_0} \right)^{\frac{1}{2}}$$
and it  is  invertible. \\

\section{Summary}
\label{s:Summ}
\noindent We have   shown a relation between   Fick-Jacobs and   Schr\"{o}dinger equation.
When the diffusion coefficient is constant, we have obtained exact solutions for Fick-Jacobs equation for different geometries. Moreover, the general case was studied
and it has shown that the solutions of Fick-Jacobs equation can be seen as a quantum state.

%\section*{ACKNOWLEDGMENTS}
%This work was partially supported (JMR) by Grant: PROMEP  No. 47510205.

\end{document}